\documentclass[conference]{IEEEtran}
\IEEEoverridecommandlockouts
\usepackage{cite}
\usepackage{amsmath,amssymb,amsfonts}
\usepackage[T1]{fontenc}
\usepackage{graphicx}
\usepackage{paralist}
\usepackage{mathbbol}
\usepackage{todonotes}
\usepackage{subcaption}
\usepackage{stmaryrd}

\usepackage{algpseudocode}
\usepackage{algorithm}
\usepackage{varwidth}
\usepackage{pdflscape}
\usepackage{float}
\usepackage{pifont}

\usepackage{paralist}
\usepackage{cancel}
\usepackage{tikz}
\usetikzlibrary{decorations.pathreplacing,positioning,matrix,backgrounds,fit,petri}
\usetikzlibrary{shapes.geometric,shadows.blur, shapes.arrows}

\usepackage{hyperref}
\newcommand{\PARstart}[1]{#1}
\newcommand{\email}[1]{\href{mailto:#1}{#1}}

\makeatletter
\newcommand{\linebreakand}{%
  \end{@IEEEauthorhalign}
  \hfill\mbox{}\par
  \mbox{}\hfill\begin{@IEEEauthorhalign}
}
\makeatother

\newtheorem{definition}{Definition}

\newenvironment{keywords}
{\par\medskip\small\textbf{\textit{Keywords:}}\enspace}
{\par\medskip}

\begin{document}
% \linenumbers

\title{A Privacy-Preserving Approach to \\ Conformance Checking}

\author{
\IEEEauthorblockN{Luis Rodr\'{\i}guez-Flores}
\IEEEauthorblockA{
    % \textit{dept. name of organization (of Aff.)} \\
    % \textit{Tecnologico de Monterrey}\\
    Tecnologico de Monterrey, Mexico \\
    \email{lrodriguez@tec.mx}}
\and
\IEEEauthorblockN{Luciano Garc\'{\i}a-Ba\~nuelos}
\IEEEauthorblockA{    % \textit{dept. name of organization (of Aff.)} \\
% \textit{Tecnologico de Monterrey}\\
Tecnologico de Monterrey, Mexico \\
\email{luciano.garcia@tec.mx}}
\linebreakand
\IEEEauthorblockN{Abel Armas-Cervantes}
\IEEEauthorblockA{
    % \textit{dept. name of organization (of Aff.)} \\
    % \textit{The University of Melbourne}\\
    The University of Melbourne, Australia \\
    \email{abel.armas@unimelb.edu.au}}
\and
\IEEEauthorblockN{Astrid Rivera-Partida}
\IEEEauthorblockA{    % \textit{dept. name of organization (of Aff.)} \\
% \textit{Tecnologico de Monterrey}\\
Tecnologico de Monterrey, Mexico \\
\email{a01324504@tec.mx}}

}

\maketitle
\thispagestyle{plain}
\pagestyle{plain}

\begin{abstract}
	Conformance checking, one of the main process mining operations, 
	aims to identify discrepancies between a process model and an event log. 
	The model represents the expected behaviour, whereas the event log represents 
	the actual process behaviour as captured in information systems' records. 
	Traditionally, the process model and the event log are both accessible to
	the business analysist performing the conformance checking. 
	However, in some contexts it is necessary to keep either the model or
	the log private to protect critical or sensitive information. In this paper, 
	we propose a secure approach to conformance checking based on 
	string processing algorithms and homomorphic encryption, where the process model and 
	event log are not visible to either the model's or event log's owner. The 
	proposed technique is based on alignments, a well-known 
	formalism used for conformance checking. An evaluation is performed using 
	a synthetic and a real-world event log, showing that conformance checking 
	can be securely computed at the expense of high memory and processing 
	requirements.	
\end{abstract}
	  
	  
\section{Introduction}

\PARstart{P}{rocess} mining is a family of tools and techniques that assists in discovering fact-based opportunities for process improvements. This is done by analysing historical process executions (event logs) recorded by information systems. Event logs record process executions as traces, which are sequences of activity instances (events) ordered by occurrence. There are four main operations in process mining: model discovery, conformance checking, variants analysis and process enhancement. 
Conformance checking, the operation relevant to this paper, compares a model -- capturing expected process behaviour -- and an event log -- observed process behaviour -- to find similarities and deviations.

There are various techniques for conformance checking, but trace alignments~\cite{AdriansyahAB2012} remain the most popular. In this technique, every trace in the event log is matched to the most similar process model execution (run). Optimal alignments describe the minimum number of necessary editions to the trace and the run to find the largest common behaviour between them. Alignments use three types of operations over event-activity pairs: \emph{match} a trace event to a model activity, \emph{move on model} to hide an activity in the model that cannot be matched to a trace event, and \emph{move on log} to hide a trace event that cannot be matched to a model activity. Usually, moves on model and on log carry a cost, hence when looking for optimal alignments the sum of the moves on model and log are minimised. 

Traditionally, the model and the event log are known to the end-user performing the conformance checking operation. However, in some scenarios the process model and event log must be kept private.
For instance, in custom manufacturing,
manufacturers strive to keep their production details secret, especially those related to the production of custom pieces made for customers developing new high tech products. Still customers
would be willing to check the progress on the production of their pieces. The latter calls for a conformance checking technique over encrypted data.
The community of process mining is already addressing some of the problems associated with privacy and confidentiality (e.g., \cite{GamalFP2021}). However, at the time of this writing, we are not aware of any other work applying homomorphic encryption to conformance checking.

In this paper, we propose a Secure Conformance Checking technique, which is outlined in Fig.~\ref{fig:approach}. 
Using public key cryptography, our approach guarantees that an entity responsible of checking the conformance of a trace with a model can do it without revealing the trace to the process model's owner.
In the presented technique, fully homomorphic encryption\footnote{Fully homomorphic encryption is a encryption scheme where additions and products can be done over encrypted data} is used to preserve confidentiality, and string processing algorithms are used to check conformance. These string processing algorithms provide efficient ways to find queries (traces) within text (model behaviour). Specifically, the FM-Index~\cite{FerraginaM00} is computed on the text representing the model behaviour. 
In order to leverage such techniques, the model behaviour must be represented as a string, hence we rely on Petri net unfolding techniques to compute complete finite representations, i.e., runs, thereof. While the presented technique is based on the computation of alignments, 
it currently supports only moves on log.
Our technique uses homomorphic operations (addition and product) to traverse the FM-Index 
to implement a secure conformance checking method.

The rest of the paper is organized as follows. Section~\ref{sec:background} introduces relevant definitions and notation.
Section~\ref{sec:priv:prev:cc} presents the main contributions of this paper, the proposed Secure Conformance Checking technique, which is evaluated in Section~\ref{sec:evaluation}.
Finally, Section~\ref{sec:conclusion} summarises the paper and gives directions for future work.

% !TEX root = ../main.tex
\section{Background}
\label{sec:background}

This section establishes the background of the paper. Subsection~\ref{subsec:nets} presents Petri nets, the modelling language adopted to represent the processes. Subsection~\ref{sub:sec:logs} presents event logs. Definitions relevant to alignments are given in Subsection~\ref{sub:sec:conf_check}. Finally, concepts related to string processing algorithms are covered in the last three subsections, FM-Index (Subsections~\ref{sub:sec:fm:index}), Backward search in the FM-Index (Subsection{sub:sec:backwards}), and LF mapping over the wavelet matrix (Subsection~\ref{sub:sec:wavelent:tree}).

\subsection{Petri nets}
\label{subsec:nets}

Petri nets is a formal model for concurrent systems. This formalism offers an intuitive graphical representation and a precise mathematical definition. 
%Furthermore, it also provides a formally defined semantics and the availability of several analysis techniques. 
In the following, we assume the reader to be familiar with Petri nets, an introduction to this formalism can be found in other works, such as~\cite{murata_1989}. Let us recap only the basics on Petri nets. 

\begin{figure}
	\centering
	\includegraphics[scale=0.5]{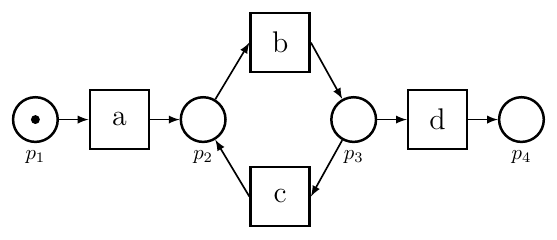}
	\caption{Petri net $N$}
	\label{fig:pn_01}
\end{figure}

A Petri net has transitions, places and arcs. Transitions represent activities or actions and are graphically represented with boxes. Places represent the state of a process and are graphically represented as circles. Arcs, represented by arrows, are used to establish an order between places and transitions. Arcs connect places to transitions, or transitions to places. Figure~\ref{fig:pn_01} shows a Petri net $N$, where transitions carry a label -- name of the activity it represents -- but transitions can be also silent (labelled as $\tau$). %Such Petri nets where transitions carry a label are called \emph{Labelled Petri nets}. 
The black dot in $p_1$ is a token, which represents resources necessary to execute an activity (\emph{fire a transition}). A distribution of tokens in a Petri net is a \emph{marking}. A transition $t$ is \emph{enabled} and can \emph{fire} if every place in its preset contains a token. Firing $t$ leads to a new marking, where a token is removed from each place in the preset of $t$ and a token is added to each place in the postset of $t$. 
While places can contain any number of tokens, in this work we consider only \emph{safe} Petri nets, i.e., every place can contain at most one token. Petri nets can have an initial and a final marking, where an initial marking represents the initial token distribution from which transitions can be fired, and the final marking is the marking from which no transition is enabled. In Fig.~\ref{fig:pn_01}, the initial marking is $\{p_1\}$ and the final marking is $\{p_4\}$. 
When transitions fire successively from an initial marking, they form a firing sequence and if it reaches the final marking, it is called a \emph{run}. For instance, two firing sequences that are also runs in $N$ (Fig.~\ref{fig:pn_01}) are $\langle a,b,d\rangle$ and $\langle a,b,c,b,d\rangle$.

In the presence of loops, a Petri net represents infinite behaviour. Complete unfolding prefixes~\cite{mcmillan1993symbolic,MontariR1994} are finite prefixes representing the partial order semantics of a Petri net. These prefixes are acyclic and contain only forward conflict (places can have only one incoming arc, but they can have several outgoing arcs). Due to this restriction, the complete prefix of a Petri net requires having several transitions and places instantiations,  which are called events and conditions, respectively. There are various strategies to construct a prefix unfolding (e.g., \cite{EsparzaRV02}). %, complete with respect to some notion of completeness. %For instance,~\cite{EsparzaRV02} presents strategies to compute prefixes complete with respect to markings. 
In this work, we use~\cite{ArmasCervantesBGD2016} to compute a prefix unfolding complete with respect to causal relations. I.e., an unfolding stops once a marking already contained in the prefix is found, and the activities executed to reach such marking is the same. We refer to this last unfolding technique as \emph{causally complete prefix unfolding}. 
 
\begin{figure}
	\centering
	\includegraphics[scale=0.5]{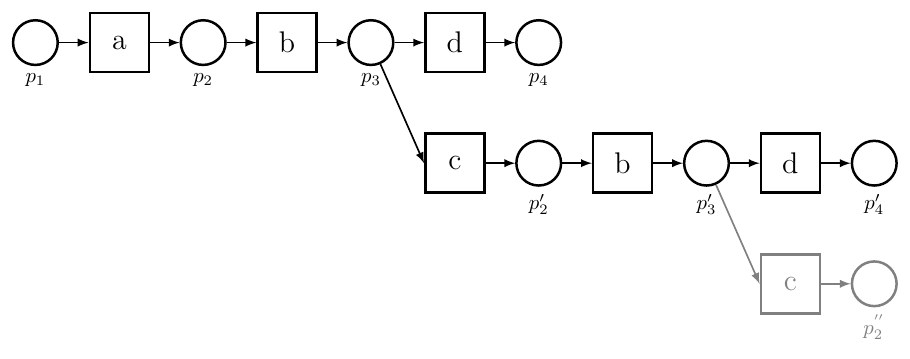}
	\caption{Causally complete prefix unfolding of $N$ (Fig.~\ref{fig:pn_01})}
	\label{fig:pn:unf_01}
\end{figure}

For example, Fig.~\ref{fig:pn:unf_01} shows the causally complete prefix unfolding of $N$ (Fig.~\ref{fig:pn_01}). In this unfolding, $p_2''$ is called the cut-off condition (a marking in the Petri net for which an equivalent marking already exists in the prefix). In this example, $p_2''$ is cut-off because $p_2'$ represents the same marking and both markings are reached through the execution of activities $\{a,b,c\}$. Please note that $p_2$ is also the same marking as $p_2'$ and $p_2''$, however $p_2$ is caused by $\{a,b\}$, and $p_2'$ and $p_2''$ are caused by $\{a,b,c\}$. 
Please note that a causally complete prefix unfolding contains complete and incomplete runs. In the example shown in Fig.~\ref{fig:pn:unf_01}, the complete runs are $\langle a,b,d\rangle$ and $\langle a,b,c,b,d \rangle$ because they reach the final marking represented by places $p_4$ and $p_4'$.

\subsection{Event logs}
\label{sub:sec:logs}

Event logs record historical business process executions. A process execution is captured in the log by means of a trace, which is a sequences of events ordered by their time of occurrence. A pair of traces are considered as the same \emph{trace variant} if they have the same number of events, which are instances of the same activities, and they were observed in the same order of occurrence.
A log $L$ can contain several occurrences of the same traces, thus a log is defined as a multiset of traces. 

Let $\mathcal{A}$ be the universe of activity labels. Then, an event is represented as a tuple $e = (t, a, c)$ where $t$ is the timestamp, $a \in \mathcal{A}$ is the activity label, and $c$ is the case id. A trace is a non-empty ordered sequence of events $\sigma = e_1, \dots, e_n$  such that $\forall i \in [1 \dots n-1]  e_i.c = e_{i+1}.c ~\land~ e_i.t \leq e_{i+1}.t$. An event log is a multiset  $L$ of traces.

\subsection{Conformance Checking}
\label{sub:sec:conf_check}

Conformance checking techniques provide mechanisms to relate modelled and observed behaviour~\cite{proc_min_handbook}.
There are three algorithmic perspectives to relate the process model and event log~\cite{CarmonaDSW18}:
rule checking, token-replay and alignments.
In this work we consider \emph{alignments}.

An alignment \cite{AdriansyahAB2012} is defined as a pairwise comparison
between a trace of the event log against an execution of the process model (\emph{a.k.a.} run). 
An alignment is composed by \textit{moves}, which are event-activity pairs $(e, a)$. 
Sometimes, deviations occurs and events cannot be matched to any activities allowed
 by the model or vice versa.
 In such cases, it is necessary to make either a model or a log move to keep matching the rest of the events and activities. %we match the event with \textit{null} activity ($\gg$).
The three types of moves in an alignment are defined as follows:
%Carmona \cite{Carmona2022ConformanceFMC} considers three types of moves: 
\begin{itemize}
    \item \textit{Synchronous move:} an event is matched with a corresponding
    activity (transition) in the process model. %It holds $e_i = a_i$ and $e_i \neq \gg$. 
    \item \textit{Log move:} an event in a trace indicates the execution
    of an activity, yet it cannot be executed in the model.
    %It holds $e_i \neq \gg$ and $a_i = \gg$.
    \item \textit{Model move:} an activity in the process model
    cannot be related to an event in a given trace. %It holds $e_i = \gg$ and $a_i \neq \gg$.
\end{itemize}

In conformance checking, the alignments of interest are \textit{optimal} -- those with the least number of log and model moves. In the reminder of the paper, when we talk about alignments we refer to optimal alignments. In this work, we only consider synchronous moves and moves on log, the computation of alignments with moves on model will be left for future work. The formal definition for alignments is presented next.

\begin{definition}[Trace alignments]
Let $\xi$ be the universe of all events, $\sigma$ be a trace and $\gamma$ be a run of a Petri net. An alignment $\psi$ is a sequence of \emph{moves} relating the events in $\sigma$ with the events in $\gamma$. Each move is a pair $(e_i, a_i)$, where $e_i \in \xi \cup \{\gg\}$ and $a_i \in \mathcal{A} \cup \{\gg\}$, such that $\gg$ denotes a missing counterpart. A move $(e_i,a_i)$ is a synchronous move if $e_i.a = a_i$, a model move if $e_i = \gg$ and $a_i \in \mathcal{A}$, or a log move if $e_i \in \sigma$ and $a_i = \gg$.
\end{definition}

\subsection{The FM-Index}
\label{sub:sec:fm:index}

In this work, string processing algorithms are used to test the conformance of a trace in a model. In particular, we use FM-Index~\cite{FerraginaM00}, which makes it possible to search a substring of length $m$ in a text string in $O(m)$ time. The FM-Index can be implemented in several ways and requires, either conceptually or concretely, the following data structures: a suffix array (SA), a Burrows-Wheeler transform (BWT), and a wavelet tree.

\begin{figure}[htb]
	% \centering
	\begin{subfigure}{0.2\textwidth}
		\centering
		\includegraphics[scale=0.4,trim={0 0 15cm 0},clip]{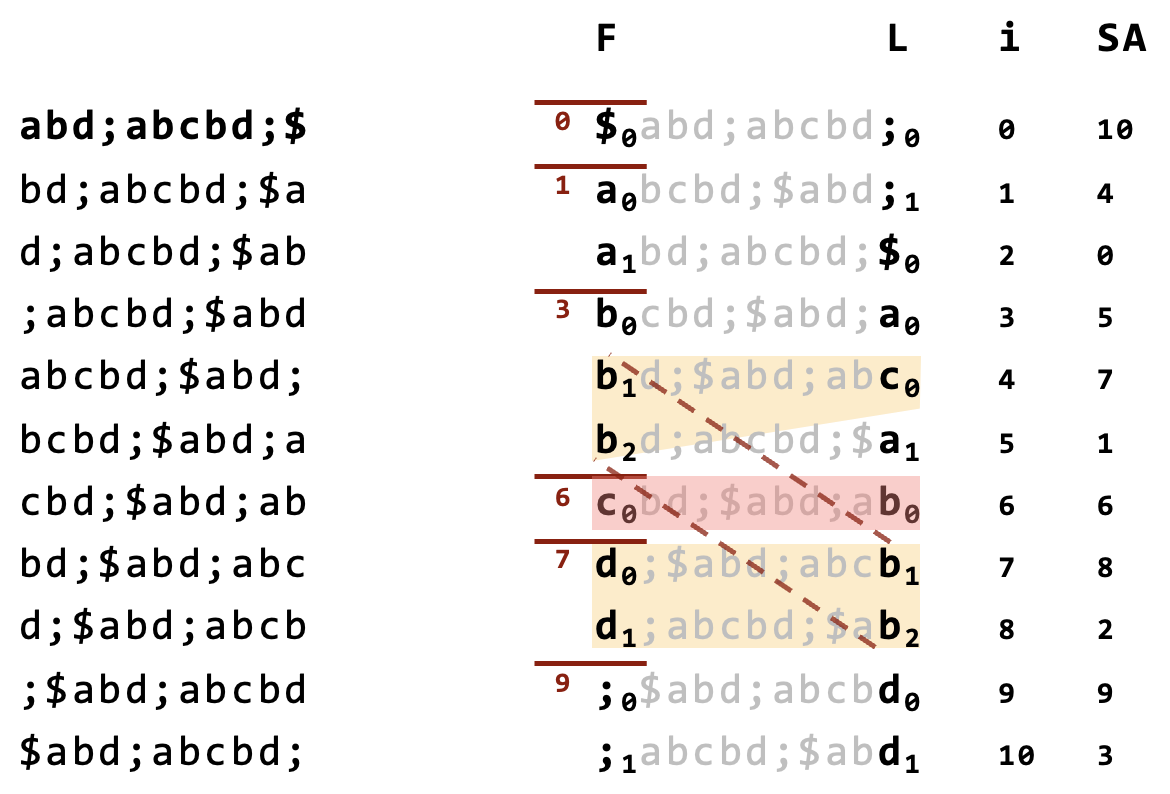}
		\caption{\label{fig:bwsearch:a}}
	\end{subfigure}
	\begin{subfigure}{0.24\textwidth}
		\centering
		% \begin{tikzpicture}
		% 	\node (m) at (0,0) {
				\includegraphics[scale=0.4,trim={9cm 0 0 0},clip]{bwt_bwsearch.png}
			% };
		% 	\draw [decorate, decoration={brace}, thick] (m.north west) -- node [above] {$\mathcal M$} ++(3.2,0);
		% \end{tikzpicture}
		\caption{\label{fig:bwsearch:b}}
	\end{subfigure}
	\caption{\label{fig:bwsearch}Computation of the BWT for {\sf abd;abcbcd;\$} (a) Matrix of circular shifts (b) Sorted matrix}
\end{figure}

In \cite{BWT_1994}, Burrows and Wheeler described a text transformation (BWT), as part of a technique for data compression. Using a permutation that is reversible, the BWT rearranges characters in a text string to achieve better compression ratios in comparison to using the original string. 
In the following, we assume that the input text is drawn from
an alphabet $\Sigma$, which includes the sentinel character \$. 
This special character is the smallest in the alphabet, is used to mark the end of a text, and never appears in the text.

The procedure to compute the BWT for a text T= {\sf abd;abcbd;\$} is illustrated in Fig.~\ref{fig:bwsearch}. 
%Let us call $T$ to the input string. 
A left circular shift of a string $s$ is another string resulting 
from removing the first character in $s$ and moving it 
back to the end. For instance, {\sf bd;abcbd;\$a} is the first 
circular shift of {\sf abd;abcbd;\$}. All the circular shifts 
of the input string are computed and collected in a matrix as 
shown in Subfig.~\ref{fig:bwsearch:a}. We proceed by sorting 
lexicographically all the circular shifts to form a matrix $\mathcal M$, 
and which is shown in Subfig.~\ref{fig:bwsearch:b}. 
From $\mathcal M$, we keep only the first and last columns, 
i.e., $F$ and $L$ respectively. The other columns can be discarded. 
By transposing column $L$ we obtain the BWT of $T$, which corresponds
to {\sf ;;\$acabbbdd} in our example. Since the BWT is also a string, 
we oftentimes refer to its characters by their indexes.

The suffix array (SA) is another text index intimately related
to the BWT. The SA is an array of integers storing the positions 
of all sorted suffixes of $T$.
Subfigure~\ref{fig:bwsearch:b} shows the SA of the string 
{\sf abd;abcbd;\$} in the last column. 
We can easily check in Subfig.~\ref{fig:bwsearch:b} that 
$T[10\!:]$\footnote{
With abuse of notation, we will use $T[4\!:]$ to denote the suffix
of $T$ starting at position 4, $T[2\!:\!4]$ to denote the substring
between positions 2 and 4, and $T[:\!4]$ to refer to the prefix
of $T$ up to position 4.
} is
the smallest lexicographical suffix of $T$ ({\sf \$}), followed by $T[4\!:]$ ({\sf a\$}), 
so on and so forth. Several linear time algorithms exist 
in the literature (e.g., \cite{NongZC11}) to compute the SA of a string. 
For a text $T$, the BWT can be computed using the SA as follows:

\begin{equation}
	BWT[i] =  
		\begin{cases}
			T[SA[i] - 1] & \text{if SA}[i] > 0\\
			\$ & \text{otherwise}.
		\end{cases}
\end{equation}

\subsection{Backward search in the FM-Index}
\label{sub:sec:backwards}

We now describe how to search for a substring in the
FM-Index. To illustrate the procedure, we will search
{\sf acbd} in {\sf abd;abcbcd;\$} using the information depicted 
in Fig.~\ref{fig:bwsearch}. This is illustrated with the highlighted area in Fig.~\ref{fig:bwsearch:b}. 
The search occurs backwards, such that the first symbol to query is {\sf d}. It starts in column
$F$ in the Burrows-Wheeler matrix (BWM) $\mathcal M$, where we can see that {\sf d} is 
in the rows 7 and 8 (see column $i$ next to $\mathcal M$ in 
Fig.~\ref{fig:bwsearch}). 
Since we can have multiple rows, we use
the interval notation. 
Therefore, we will say that {\sf d} occurs in the interval
$\llbracket 7,8\rrbracket$\footnote{Please note that, in order to avoid confusion with the references, we use $\llbracket\ \rrbracket$ and $\llparenthesis\ \rrparenthesis$ to represent closed and open intervals with the usual meaning.} of $F$. We proceed with the second symbol, {\sf b}. However,
now we will look for the occurences of {\sf b} in the inverval $\llbracket 7,8\rrbracket$
over column $L$. 
%Rows 7 and 8 in $\mathcal M$ are highlighted in Fig.~\ref{fig:bwsearch}, to help the reader follow the steps in this procedure.
%
{\sf d} is always preceeded by {\sf b} in the indexed string.
That is why 2 {\sf b}s are selected in $L$, leaving out of the 
analysis one additional {\sf b}. Please note that we added a subscript to
all the letters, which is referred to as the rank of the letter. 
The latter means that "$b_1$" and
"$b_2$" precede "$d_0$" and "$d_1$" in the indexed string, respectively.
So, now our search continues in the block for {\sf b}s that appear
over column $F$ including only the {\sf b}s with ranks 1 and 2, 
which appear in the interval $\llbracket 4,5\rrbracket$ in $F$.
Note that letters are grouped together in $F$ forming blocks, because
the suffixes are alphabetically sorted.
Once again, we proceed by looking for occurrences of {\sf c}s in the
interval $\llbracket 4,5\rrbracket$ over $L$. In this case, there is only {\sf c}s in that
interval, i.e., {\sf c\ $\!_0$}.
We search {\sf c\ $\!_0$} in $F$ and
find it in the interval $\llbracket 6,6\rrbracket$. Then, we have to look for {\sf a} over
interval $\llbracket 6,6\rrbracket$ in $L$. However, there is no {\sf a} in such interval, 
implying that {\sf acbd} is not a substring of {\sf abd;abcbcd;\$}.

The above procedure assumes that both $F$ and $L$ are stored and that
$F$ is traversed every time we determine the interval associated with
a given pair of characters (with different or the same ranks). Instead, 
precomputed information can be stored in auxiliary data structures to
enable quick look up, by implementing the following functions:

\begin{itemize}
\item {\sc Rank}($T, c, i$) returns the number of occurrences of character \emph{c}
	in the prefix $T[ :\!i]$,
\item {\sc C}($T, c$) returns the overall number of characters in $T$
	that are strictly lexicographically smaller than character \emph{c}.
\end{itemize}

For convenience, the function
{\sc LF}($T, c, i$) $\triangleq$ {\sc C}($T, c$) + {\sc Rank}($T, c, i$)
is used, where "LF" stands for "Last to First mapping".
If function {\sc LF} is available, the overall backwards search can be 
implemented as shown in Algorithm~\ref{algo:bwsearch}.

\begin{algorithm}
	\begin{algorithmic}[1]
\Function{BackwardSearch}{BWT, q, i, j}
	\For {c $\in$ \Call{Reverse}{q} {\bf while} i $<$ j}
		\State i, j $\gets$ \Call{LF}{BWT, c, i}, \Call{LF}{BWT, c, j}
	\EndFor
	\State \Return $\llbracket\text{i, j}\rrparenthesis$
\EndFunction
	\end{algorithmic}
\caption{\label{algo:bwsearch} Backward search over Burrows-Wheeler transform}
\end{algorithm}

\section{Algorithms}

\subsection{LF mapping over the wavelet matrix}
\label{sub:sec:wavelent:tree}

\begin{figure}[h]
    \flushleft
    \includegraphics[scale=0.65]{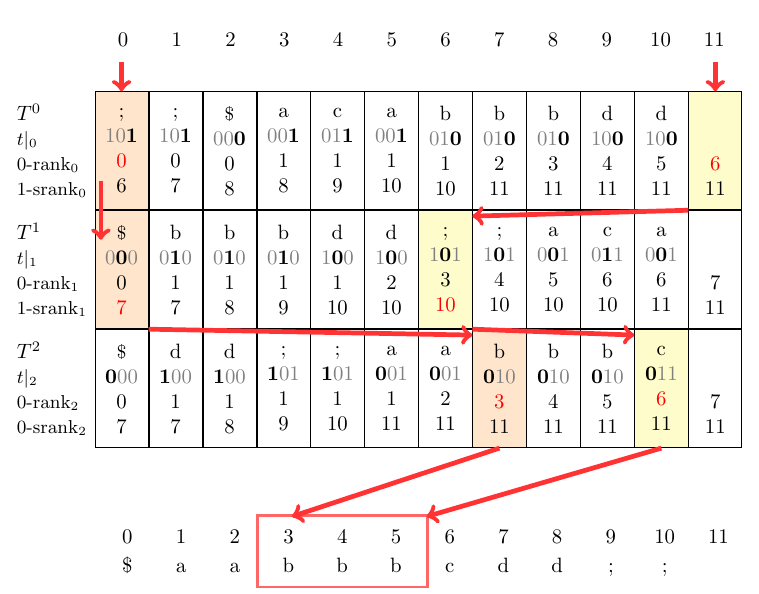};
	\caption{\label{fig:wmatrix}Wavelet matrix}
\end{figure}

Several data structures can be used to implement the
computation of {\sc C} and {\sc rank} functions.
We adopt the wavelet matrix (WM) presented in \cite{SudoJNS19} and
extend the protocol defined thereof, to define our secure conformance
checking method.
Conceptually, the WM can be seen as a matrix that tracks the passes of the
radix sort starting with the $L$ column of the BWM and finishes with
the column $F$ of the BWM. The WM of our running example is presented 
in Fig.~\ref{fig:wmatrix}.

The WM captures the passes of the radix sort, more specifically
a bitwise radix sort. Hence, we need some additional notation. Let us consider
an alphabet $\Sigma$, from which we take the characters that compose the
indexed strings. For convenience, each character is encoded using a value in
$\llbracket 0, |\Sigma|\rrbracket$ and, hence, the WM has $\text{log}_2 |\Sigma|$ rows.
For our running example, we consider the encoding $\{(\text{\sf \$}, 0), (\text{\sf a}, 1),
(\text{\sf b}, 2), (\text{\sf c}, 3), (\text{\sf d}, 4), (\text{\sf ;}, 5)\}$, which
implies a 3 bit long encoding. Let $\alpha|_i$ denote the i-th bit in the code associated 
with character $\alpha$ taken from alphabet $\Sigma$. Now, the first row in the sWM corresponds
with the BWT's $L$ column, as shown in Fig.~\ref{fig:wmatrix}. The second row rearranges
the characters from the first row, according to the least-significant bit on the encoding for
each character. In this way, the occurrences of {\sf b} in the matrix go before the occurrences
of {\sf a}, because $\text{\sf b}|_0 = 0$ and $\text{\sf a}|_0 = 1$. However, there is
a {\sf c} between two {\sf a}s in the first row and their order is preserved in the second row.
This property, known as stability, is important because the radix sort preserves the rank of
the characters at every iteration. Similarly, the order of the characters in the third and forth
row (shown separately in the figure), are based on the second and third bits, respectively.
Note that the characters in the forth row are lexicographically sorted, as is the case for
BWM's $L$ column.

Each cell in the WM carries two additional values, namely the bitwise 0-rank 
($\text{\sf 0-rank}_r$, with $r$ refering to the row number) and the bitwise shifted
1-rank ($\text{\sf 1-srank}_r$). They can be formally defined as:

% \LN{Podemos reutilizar las deficiones previas para definir 0-rank y 1-rank (como en el paper original rankCF).
% 	\begin{itemize}
% 		\item {\sc Rank}($t|r, val, i$) that returns the number of occurrences of \emph{val} in the prefix $t|r\llbracket :\!i - 1\rrbracket$,
% 		\item {\sc C}($t|r, val$) that returns the overall number of values in $t|r$ that are strict smaller than \emph{val}.
% 	\end{itemize}

% 		{\sc LF}($t|r$, val, i) = {\sc C}($t|r$, val) + {\sc Rank}($t|r$, val, i)  \\

% 		\vspace{0.5cm}

% 		Lo que usted tiene definido como 0-rank o 1-rank serian:\\ \vspace{0.2cm}

% 		0-rank es: {\sc LF}($t|r$, 0, i) = {\sc C}($t|r$, 0) + {\sc Rank}($t|r$, 0, i)  \\
% 		1-rank es: {\sc LF}($t|r$, 1, i) = {\sc C}($t|r$, 1) + {\sc Rank}($t|r$, 1, i)  \\

% }

\begin{small}
\begin{equation*}
    \text{\sf 0-rank}_r(T, p) = \sum_{i = 0}^{p-1} \begin{cases}
        1, & \text{if $T[i]|_r$ = 0}\\
        0, & \text{otherwise}
     \end{cases}
\end{equation*}
\end{small}

\begin{small}
\begin{equation*}
        \text{\sf 1-srank}_r(T, p) = \sum_{i = 0}^{|T|} \begin{cases}
        1, & \text{if $T[i]|_r$ = 0}\\
        0, & \text{otherwise}
     \end{cases} + \sum_{i = 0}^{p-1} \begin{cases}
        1, & \text{if $T[i]|_r$ = 1}\\
        0, & \text{otherwise}
     \end{cases}
\end{equation*}
\end{small}

% \begin{equation*}
%     \text{\sf 0-rank}_r(T, p) = |\{ i\ |\ 0 \le i < p \land T[i]|_r = 0 \}|
% \end{equation*}
% \begin{equation*}
%     \text{\sf 1-srank}_r(T, p) = |\{ i\ |\ 0 \le i < |T| \land T[i]|_r = 0\}| + |\{ i\ |\ 0 \le i < |T| \land T[i]|_r = 1 \}|
% \end{equation*}

$\text{\sf 0-rank}_r$ is the rank of each character, with 0 at position 
$r$ on the binary encoding associated with the character. 
In other words, $\text{\sf 0-rank}_r$ counts the number 
of characters that have a bit 0 in string $T$ up to a given position. 
Similarly, $\text{\sf 1-srank}_r$ computes the rank of each character 
in the input string, with 1 in the $r$-th position. However, the 
$\text{\sf 1-srank}_r$ uses an initial offeset that corresponds to the 
count of all characters in string $T$ having 0 at position $r$ on the 
binary encoding associated thereof. Instead of computing dynamically the
values of $\text{\sf 0-rank}_r$ and $\text{\sf 1-srank}_r$, their values
are stored in the WF and accessed using conceptual matrices,
i.e., {\sc 0-rank}[{\sf r, i}] and {\sc 1-srank}[{\sf r, i}] where
{\sf r} and {\sf i} represent the row and column indexes, respectively.

\begin{algorithm}
\begin{algorithmic}[1]
\Function{LF}{BWT, c, f, g}
    \State {\sc 0-rank}, {\sc 1-srank} $\gets$ WM(BWT) 
    % \\ \Comment Recall the matrices are precomputed, so this step is performed only once
    \For {r = 0 {\bf to} $\lceil \text{log}_2 |\Sigma| \rceil$}
        \State f $\gets$ {\sc 0-rank}[r, f] {\bf if} c|$_r$ = 0 {\bf else} {\sc 1-srank}[r, f]
        \State g $\gets$ {\sc 0-rank}[r, g] {\bf if} c|$_r$ = 0 {\bf else} {\sc 1-srank}[r, g]
    \EndFor
\State \Return $\llbracket f, g \rrparenthesis$
\EndFunction
\end{algorithmic}
\caption{\label{algo:lf}Last to First mapping}
\end{algorithm}

Let us refer to Fig.~\ref{fig:wmatrix} to illustrate how the WM is
traversed, to compute the LF mapping for character {\sf b}. In the
example, it is assumed that {\sf b} is queried in all the indexed 
string, i.e., the interval $\llbracket 0, 11 \rrparenthesis$,
such that {\sf f} and {\sf g} are initially set to 0 and 11, respectively. 
Since the alphabet for the example has 7 characters, the loop in 
lines 4-7 is repeated 3 times. In the first iteration, r gets the 
value of 0 (row 0). The code associated with {\sf b} is
{01{\bf 0}} such that b|$_0$ is 0. In Fig.~\ref{fig:wmatrix}
the values for {\sc 0-rank}[0, 0] and {\sc 0-rank}[0, 11] are
shown in red to help the reader following the example.
Therefore, the values of {\sf f} and {\sf g} are set in this
iteration (line 5, 6 in the algorithm) to 0 and 6.
In the second iteration, we have that b|$_1$ is 1 such
that we have to look up for {\sc 1-srank}[1, 0] and 
{\sc 1-srank}[1, 6]. The latter results in setting {\sf f} and 
{\sf g} to the values of 7 and 10. Finally, b|$_2$ is 0, such 
that {\sf f} and {\sf g} are set to the values of {\sc 1-srank}[2, 7] and 
{\sc 1-srank}[2, 10], which correspond to 3 and 6. 
The function returns $\llbracket 3, 6 \rrparenthesis$, which is the expected result.

\section{Approach}
\label{sec:priv:prev:cc}

\begin{figure}[h]
    \centering
    \includegraphics[scale=0.6]{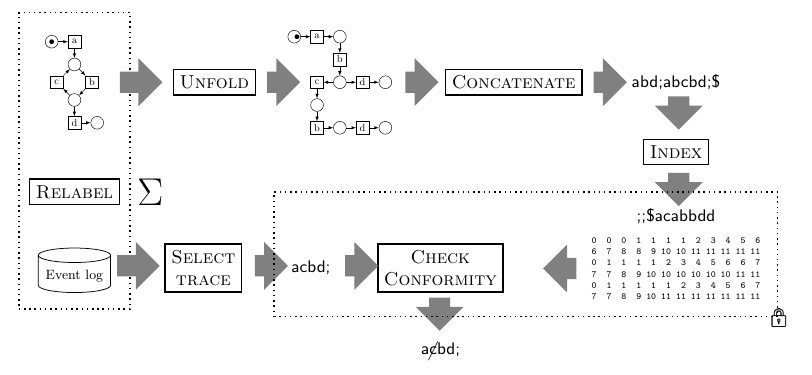}
    \caption{\label{fig:approach}Steps and datafow in our approach}
 \end{figure}

Our approach to conformance checking is mapping trace alignment 
to approximate substring matching, using a backward search 
over the BWT.
Next, each of the steps in our approach is presented (see Fig.~\ref{fig:approach}).

\subsubsection{Relabel}
It is assumed that the set of labels on both model and log coincide.
Then, the process model and event log are analyzed to determined the
set of activity/event labels, which are mapped into an
alphabet $\Sigma$ which, in turn, are mapped to values in the
range $\llbracket 0, \dots, |\Sigma|-1 \rrbracket$.
Although relabeling may happen at several points, we assume 
it happens on the input process model and when selecting
the trace to be checked for conformance.

\subsubsection{Unfold}
Process models
are assumed to be Petri nets.
The causally complete prefix unfolding~\cite{ArmasCervantesBGD2016}, 
as described in Section~\ref{sec:background}, is computed on the input Petri net 
from which only the complete executions (those reaching a final marking) 
are kept.
Runs, usually partial orders, are linearized.

\subsubsection{Concatenate}
The traces representing the complete runs in the unfolding are concatenated into a single string.
The string includes two special symbols: one marks the end of run $;$, and the other 
marks the end of the set of runs \$. The examples 
in previous sections follow this convention.

\subsubsection{Index}
The string resulting from the concatenation is indexed.
The FM-Index is built upon the BWT of the
concatenated string and its wavelet matrix computed.

\subsubsection{Select trace}

It is assumed that conformance checking is requested by the end-user,
who submits a trace to be checked. 
%Giving the performance of the methodit is likely that the conformance checking will be done only over a small set of traces, one at a time.

% ===================================================
% ===================================================
\begin{figure*}
    \centering
    \includegraphics[scale=0.8]{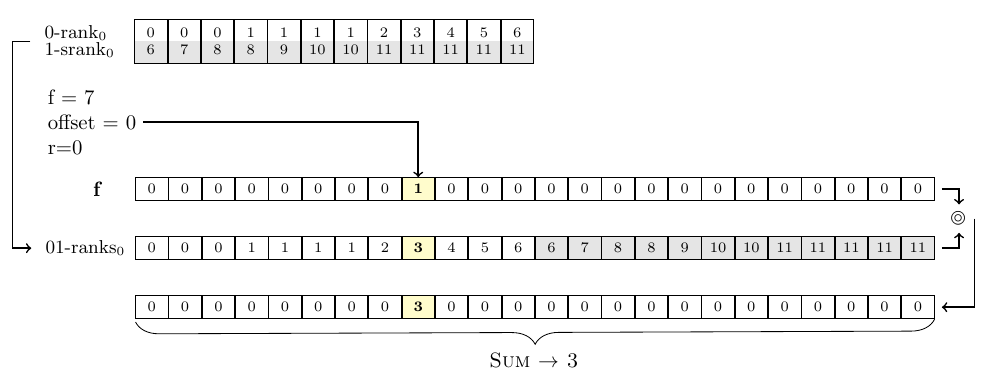}
\caption{\label{fig:vectors}Vector-oriented data flow}
\end{figure*}
% ================================================
% ================================================

\subsection{Privacy-preserving conformance checking}

Algorithm~\ref{algo:bwsearch} combined with Algorithm~\ref{algo:lf} 
computes the \emph{maximum common substring} for a query string over
the indexed string. 
The combined algorithm computes a perfectly fitting alignment for a given trace. 
However, the algorithm stops whenever a mismatch is found, such that 
it has to be extended to handle mismatches. The current approach 
allows only for log move  mismatches. 
%\todo{I think we should mention this log-move limitation earlier. Intro/background. See red text in introduction.}

Algorithm~\ref{algo:PPCCheck} outlines the privacy-preserving conformance checking
method. Before going into detail, let us summarize some general aspects about
the design of the algorithm:

\begin{itemize}
    \item The algorithm is structured to follow the Client-Server architecture. The
          client and the server are presented in boxes on the left- and on the right-hand
          side of Algorithm~\ref{algo:PPCCheck}, respectively. 

    \item The server side receives the models' runs, in the form of a concatenated
          string, which is then indexed: it computes the BWT
          and its WM.
    \item The server responds to (encrypted) queries, incrementally computing the LF mapping
          by traversing the WM.

    \item The client side receives the (query) trace to check its conformity.
        %   upon which the conformance 
        %   checking is performed. 
          It iterates over the set of event labels of the trace
          and then over the bits of the corresponding encoding, applying the backward
          search described in Subsection~\ref{sub:sec:backwards}. Intermediary results
          are obfuscated though.
    \item At the beginning of each iteration, the indexes are stored in auxiliary variables.
          Whenever a mismatch is found, the stored indexes are restored. This is the approach
          to implement the log moves.
    \item Client and server share the client's public key to enable homomorphic encryption.
          Although the client can decrypt intermediate results (using priv. key), they are obfuscated.
\end{itemize}

Let us now look into the details of the algorithm, first assuming data exchange occurs
without encryption. For simplicity, in the algorithm we do not explicitly include encryption keys, it is done in every crypto-operation.

Indexes are not sent as scalars, but instead they are packaged into vectors. Packaging the
indexes in this way is required to reduce data disclosure as it will be discussed later.
Consider the example shown in Figure~\ref{fig:vectors}. In that example, it is assumed that {\bf f} (the index)
has the value of 7, r is 0 (it is the iteration corresponding to the first bit), and offset
is 0 (the first bit has the value of 0).
Although the matrix in the running example has 12 columns, a vector of twice that size is 
allocated, which will be referred to as {\bf f}. 
All the elements in the vector are set to zero and only the element in 
position 7 is set to 1. For convenience, the {\sf 0-rank$_0$} and the {\sf 1-srank$_0$} 
rows are concatenated to form a single vector, which will be referred to as 
{\sf 01-ranks$_0$}. An element-wise product of {\bf f} and {\sf 01-ranks$_0$},
denoted $\text{\bf f} \varocircle  \text{\sf 01-ranks$_0$}$, is
performed as illustrated in Fig.~\ref{fig:vectors}. As expected, only the element 
in position 7 in the result is non zero. As one last step, all the elements 
of the resulting vector are summed up and the result, 3 in this case, is 
the value that the server returns to the client. Recall all the explanation
here assumes that data is not encrypted.

The privacy preserving version, follows the same steps as above, except that
the vector {\bf f} is encrypted.
Note that we use a randomized encryption method 
that yields different values for the input (plain) value. That means that 
the encrypted vector will carry multiple different values, making it difficult to
guess where the single 1 is located.
Moreover, the element-wise product uses an homomorphic product, and the summation of
the elements in the resulting vector uses also an homomorphic addition. Please
note that the steps correspond to a dot product of the two vectors, which
we will denote in the algorithm as $\odot$. The algorithm requires also to
add a random value to the result of the dot product. In the algorithm, we 
use $\oplus$ to denote such homomorphic addition.

\begin{algorithm*}[htb]
    \centering
    \includegraphics[scale=1.1]{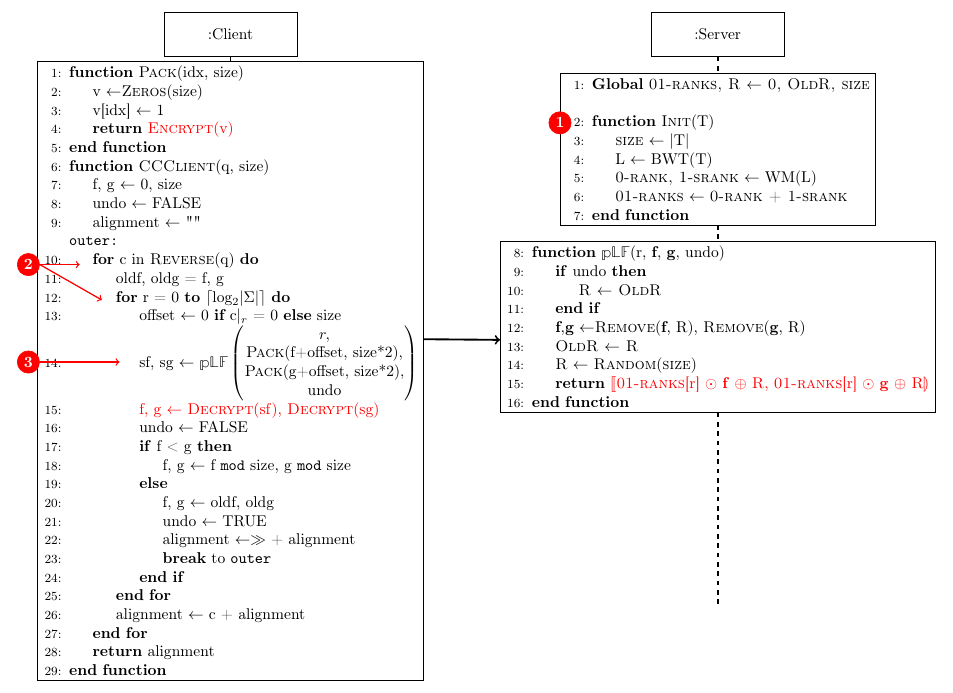}
\caption{\label{algo:PPCCheck}Privacy-preserving conformance checking}
\end{algorithm*}

Let us run through Algorithm~\ref{algo:PPCCheck}. First, 
the server side creates the FM-index. This is done in function {\sc Init} 
(see 
\textcolor{red}{\ding{182}}),
which is called with 
string T representing the concatenated model runs.
Function {\sc Init} computes the size of T and stores it in the global variable 
{\sc size}. It also computes all index data structures (BWT, WM, etc.)
and retains {\sf 01-ranks$_r$} in a global variable.

Then, a trace can be checked for conformance by calling function {\sc CCClient}
with the trace {\sf q} and the {\sf size} of the concatenated string representing 
the model runs. Note that {\sf size} is disclosed to the client.
Let us consider the flow when the trace fits one model run. You will notice
that we start the search considering the entire index (line 9, client side).
Then, the algorithm iterates over each character {\sf c} in {\sf q} backwards, and
then for each bit in the encoding for {\sf c}
(see  
\textcolor{red}{\ding{183}}).
Recall the inner loop traverses the WM to compute the LF mapping, hence the function
name $\mathbb{pLF}$, standing for partial LF. In preparation for calling $\mathbb{pLF}$ (see  
\textcolor{red}{\ding{184}}),
the indexes {\sf f} and {\sf g} are packaged into vectors (function {\sc Pack} in lines 1-5, client side).
Note that the vector is encrypted in line 4 (client side). The partial LF mapping
is computed in the server side in line 15. Please note that 
$\llbracket ${\sc 01-ranks}[r] $\odot\ \mathbf{f}\ \oplus$ {\sc R} includes a random
value to obfuscate the resulting interval, i.e., the client side is 
unable to determine the intermediate intervals during conformance checking. The
value of {\sc R} is removed at every following iteration in line 12 (server side).
Thus, $\mathbb{pLF}$ returns the obfuscated, still encrypted indexes. The client side
decrypts the indexes in line 15 (client side). A sanity check happens in line 17.
The modulus operations in line 18 are required because homomorphic operations 
may overflow; this is a common step in cryptographic protocols.
If the inner loop completes without any problem, the current character induces
a match and, hence, the character is added to the alignment variable in line 26.

To support the log move mismatches, the client side stores the indexes at the beginning
of each iteration in line 11. When a mismatch is detected, i.e., condition in line 17
is false, the indexes are restored in line 20 and the flag {\sf undo} is set to TRUE
in line 21, to restore the indexes of the previous iteration (line 11).
The flag {\sf undo} is sent to the server, to restore the {\sc R} of the
previous iteration, i.e., {\sc OldR}, in lines 9-11.
We further added the support to handle a ``budget'' for mismatches,
to cope with attacks where the client guesses the next character repeatedly
to discover a fitting trace. This feature is not captured in the Algorithm though.

It can be easily checked that the basic backwards search, resulting 
from combining Algorithm~\ref{algo:bwsearch} and Algorithm~\ref{algo:lf},
has a complexity of $O(|q| \cdot \lceil\text{log}_2 |\Sigma|\rceil )$. Conversely,
the complexity of the privacy-preserving version is dominated by the encryption
operations (redish font in Algorithm~\ref{algo:PPCCheck}). 
Let $C$ stand for a large, non negligeable constant that corresponds
to the time an homomorphic operation or the encrypt/decrypt operation take.
Hence, the encryption of each vector has a complexity of $O(2C|T|)$, i.e., encryption
of $2|T|$ values. Each one of the dot products performed in the server side 
requires $O(2C|T|)$ homomorphic products and $O(2C|T|)$ homomorphic additions,
if we also include the addition of {\sc R}. We additionally need to decrypt the
indexes, with a complexity of $O(2C)$. Therefore, the overall 
asymptotic complexity of our method is 
$O(|q| \lceil\text{log}_2 |\Sigma|\rceil C|T| )$ and, hence dominated by 
$O(C\cdot|q|\cdot|T|)$.

%\input{tex/05_CnRank_arrays}
%\input{tex/05.1_secure_CnRank_arrays}

% !TEX root = ../main.tex
\section{Evaluation}
\label{sec:evaluation}

The secure conformance checking method was implemented in C++~\footnote{https://github.com/CC\_sWM\_ROT}.
% , using the standard template library.
Homomorphic encryption was implemented using the lifted-ElGamal library~\footnote{https://github.com/herumi/mcl, v1.86.0},
with a standard elliptic curve over a 192-bit prime field.
Client-server interactions use the NNg library~\footnote{https://github.com/nanomsg/nng} and
Google's protobufs~\footnote{https://protobuf.dev}. 
% All the experiments run locally, using Unix sockets.
We run all the experiments in a Lenovo P1 laptop workstation with an Intel Core i7-10850H CPU@2.70GHz, 32GB of RAM, running Ubuntu 22.04.4 LTS.
Client and server run in the same machine using TCP sockets as the transport layer.

\subsection{Datasets}

For the evaluation, we used one public real-world event log, and two additional ones syntheticly created for their use in research in 
process discovery~\cite{LuFBA16}. The underlying process models are small to medium size and include the typical control flow patterns, i.e., sequence, blocks of paralellism, exclusive choice and loops. The characteristics of the models/event logs are summarized in Table~\ref{table:logs}.

% \begin{table}[ht]
% \begin{tabular}[ll]
% Dataset & utr \\
% \end{tabular}
% \end{table}

\subsection{Experimental design and results}

We firstly evaluated the performance of our method with Dataset G, which is a small process with two blocks of concurrency and no loops. Despite being a small model, 9 events in total, due to the presence of concurrency it has a large number of runs, namely 144 runs. Therefore, the size of the index is also larger than those of the other logs, with an impact in the execution time of the aligment of each trace. Note that all traces have the same size in this dataset. The average execution time for all traces was of 1.12 minutes with a standard deviation of 0.056. Considering that all traces have 9 events, we can say that aligning each symbol takes in avarege 0.112 minutes, if we consider that we have to add the trace separator. All traces in the log are fitting. Therefore, we selected randomly 4 pairs of traces and injected to each pair 1, 2, 3 and 4 events at random positions. We observed, however, that the average time to align each symbol remains quite similar, i.e., 0.114 minutes. We observe that a log move requires just reseting some values the server side and no additional homomorphic operations.

Figure~/ref{fig:scatter1} shows the execution times taken by our method with dataset F. The model of this dataset includes a loop. For the experiment, we decided to unroll only once the loop. Therefore, in the cases where the event logs included traces capturing looping behaviour, the second or subsequent iterations are expected to induce log moves. Note that in practice, it would expected that loops are unrolled more than once as it is typically observed in the real-world process.

\begin{figure}[htp]
	\centering
	\includegraphics[scale=0.38]{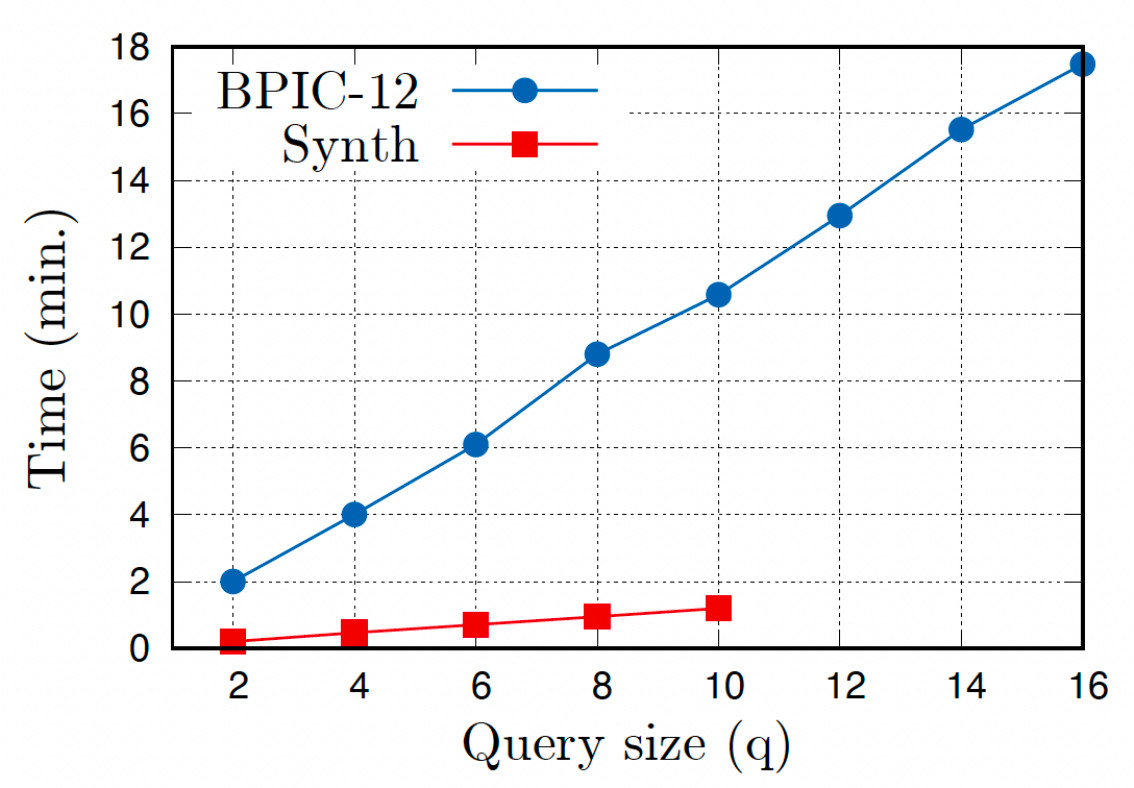}
	\includegraphics[scale=0.6]{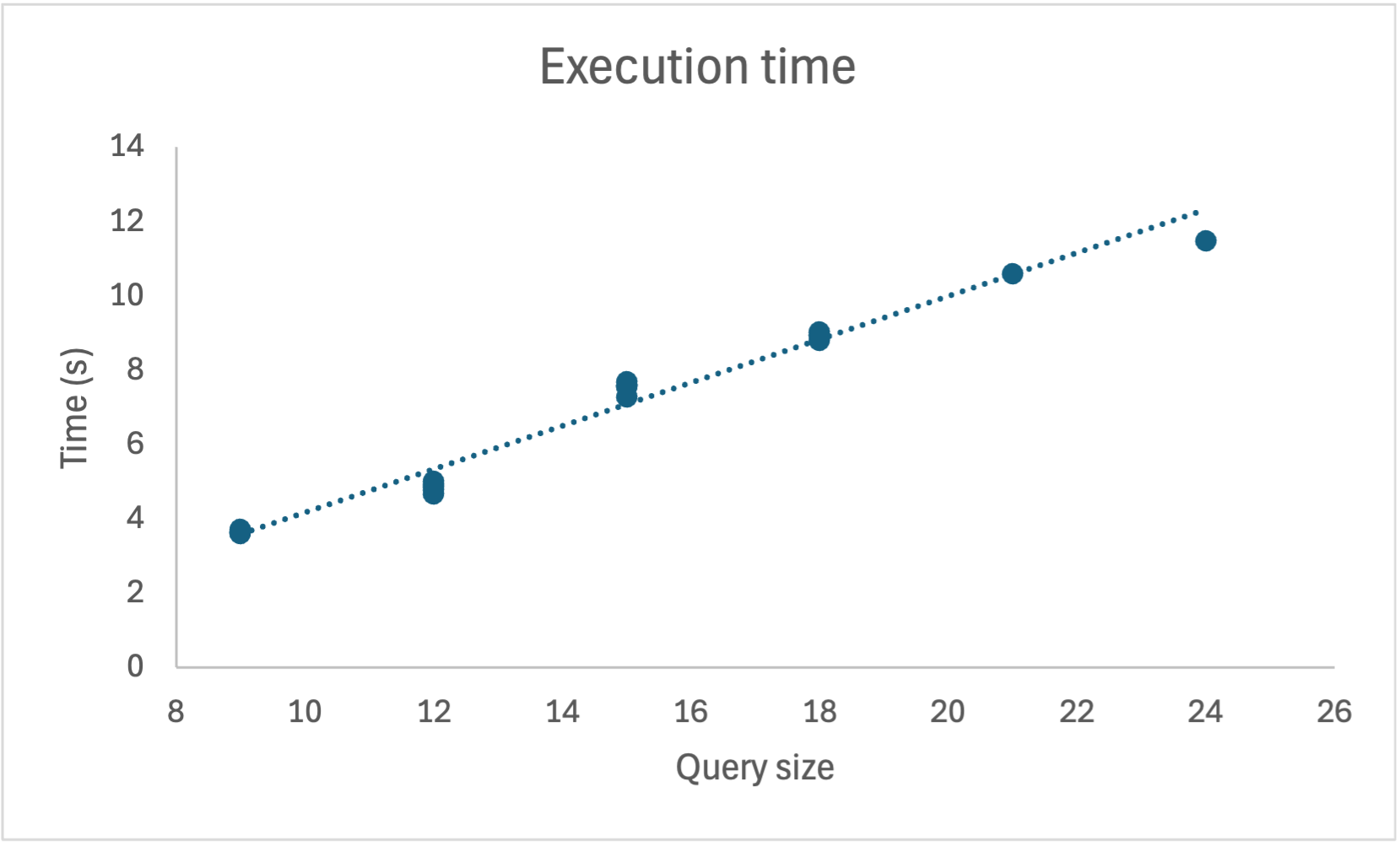}
	\caption{\label{fig:scatter1} Execution times for dataset F}
\end{figure}

We can observe a quasi-linear trend in the execution time

\section{Conclusion}
\label{sec:conclusion}

In this paper, we proposed a Secure Conformance Checking 
technique. It uses homomorphic encryption for preserve confidentiality, and string 
processing algorithms to check for conformance between a trace and a model.
The technique relies on a client-server architecture, where the client
sends a trace to check for conformance to the server, which owns the model,
 such that neither the model is visible to the client nor the trace is visible 
to the server. 
At the time of this writing, we are not aware of any other work applying 
homomorphic encryption to conformance checking.

Our conformance checking approach maps the problem of finding optimal 
trace alignments to that of substring matching, but supports only log moves.
The use of state-of-the-art string matching techniques enabled to
achieve quasi-linear complexity and to use 
homomorphic encryption. The traversal of the wavelet matrix, central to
string matching, is implemented with homomorphic products and
additions. However, the high computational cost of homomorphic operations
dominated the execution times.
As a future work, we will explore how to adapt the approach to handle 
moves on models, as well as to enhance the technique to handle
loops with multiple repetitions beyond the unfolding.

\section*{Acknowledgements}
This work has received funding from the Swiss National Science Foundation 
under Grant No. IZSTZ0 208497 (ProAmbitIon project).

\bibliographystyle{splncs04.bst}
\bibliography{bibliography.bib}

\end{document}